\documentclass[journal,comsoc]{IEEEtran}

\usepackage[T1]{fontenc}

\ifCLASSINFOpdf
\else
\fi

\usepackage{amsmath}
\usepackage[cmintegrals]{newtxmath}

\usepackage{subfigure}
\usepackage{graphicx, amssymb, amsbsy, amsmath}
\usepackage{subfigure}
\usepackage{multirow}
\usepackage{cite,url}
\usepackage{framed}
\usepackage{color,soul}
\usepackage[font=footnotesize,labelfont=sf,textfont=sf]{caption}
\usepackage{algorithm}
\usepackage{algpseudocode}
\DeclareGraphicsExtensions{.eps}
\usepackage[english]{babel}
\usepackage{blindtext}
\usepackage{epsfig,endnotes,graphicx}
\usepackage{array}
\usepackage{subfigure}
\usepackage{url}
\usepackage{amsmath}
\usepackage{fancyhdr}

\usepackage{amssymb}

\DeclareMathOperator*{\argmin}{argmin}

\def\squareforqed{\hbox{\rlap{$\sqcap$}$\sqcup$}}
\def\qed{\ifmmode\squareforqed\else{\unskip\nobreak\hfil
		\penalty50\hskip1em\null\nobreak\hfil\squareforqed
		\parfillskip=0pt\finalhyphendemerits=0\endgraf}\fi}

\ifodd 1
\newcommand{\com}[1]{\textbf{\color{red} (COMMENT: #1)}} 
\newcommand{\comg}[1]{\textbf{\color{green} (COMMENT: #1)}}
\newcommand{\response}[1]{\textbf{\color{magenta} (RESPONSE: #1)}} 
\else

\newcommand{\com}[1]{}
\newcommand{\comg}[1]{}
\newcommand{\response}[1]{}
\fi

\begin{document}

\title{Edge AI: On-Demand Accelerating Deep Neural Network Inference via Edge Computing}

\author{En~Li,
        Liekang~Zeng,
        Zhi~Zhou,~\IEEEmembership{Member,~IEEE}
        and~Xu~Chen,~\IEEEmembership{Member,~IEEE}
        \thanks{Manuscript received March 25, 2019; revised July 19, 2019; accepted September 25, 2019.}
\thanks{Part of the results of this journal version paper was presented at 2018 ACM SIGCOMM MECOMM \cite{Li2018edgeintelligence}.}
\thanks{The authors are with School of Data and Computer Science, Sun Yat-sen University, Guangzhou 510006, China (e-mail: \{lien5, zenglk3\}@mail2.sysu.edu.cn; \{zhouzhi9, chenxu35\}@mail.sysu.edu.cn).
(\textit{En Li and Liekang Zeng contribute equally to this work. Corresponding author: Xu Chen.})
}
}

\markboth{IEEE Transactions on Wireless Communications, ~Vol.~xx, No.~x, October~2019}%
{Shell \MakeLowercase{\textit{et al.}}: Edge AI: On-Demand Accelerating Deep Neural Network Inference via Edge Computing}

\maketitle

\begin{abstract}
As a key technology of enabling Artificial Intelligence (AI) applications in 5G era, Deep Neural Networks (DNNs) have quickly attracted widespread attention.
However, it is challenging to run computation-intensive DNN-based tasks on mobile devices due to the limited computation resources.
What's worse, traditional cloud-assisted DNN inference is heavily hindered by the significant wide-area network latency, leading to poor real-time performance as well as low quality of user experience.
To address these challenges, in this paper, we propose \textsf{Edgent}, a framework that leverages edge computing for DNN collaborative inference through device-edge synergy.
\textsf{Edgent} exploits two design knobs: (1) DNN partitioning that adaptively partitions computation between device and edge for purpose of coordinating the powerful cloud resource and the proximal edge resource for real-time DNN inference; (2) DNN right-sizing that further reduces computing latency via early exiting inference at an appropriate intermediate DNN layer.
In addition, considering the potential network fluctuation in real-world deployment, \textsf{Edgent} is properly design to specialize for both static and dynamic network environment.
Specifically, in a static environment where the bandwidth changes slowly, \textsf{Edgent} derives the best configurations with the assist of regression-based prediction models, while in a dynamic environment where the bandwidth varies dramatically, \textsf{Edgent} generates the best execution plan through the online change point detection algorithm that maps the current bandwidth state to the optimal configuration.
We implement \textsf{Edgent} prototype based on the Raspberry Pi and the desktop PC and the extensive experimental evaluations demonstrate \textsf{Edgent}'s effectiveness in enabling on-demand low-latency edge intelligence.
\end{abstract}


\IEEEpeerreviewmaketitle

\section{Introduction}

\IEEEPARstart{A}{s} the backbone technology supporting modern intelligent mobile applications, Deep Neural Networks (DNNs)  represent the most commonly adopted machine learning technique and have become increasingly popular.
Benefited by the superior performance in feature extraction,
DNN have witnessed widespread success in domains from computer vision \cite{szegedy2015going}, speech recognition \cite{Oord2016WaveNet} to natural language processing \cite{Wang2015LSTM} and big data analysis\cite{almomani2018online}.
Unfortunately, today's mobile devices generally fail to well support these DNN-based applications due to the tremendous amount of computation required by DNN-based applications.

In response to the excessive resource demand of DNNs, traditional wisdom resorts to the powerful cloud datacenter for intensive DNN computation.
In this case, the input data generated from mobile devices are sent to the remote cloud datacenter, and the devices receive the execution result as the computation finishes.
However, with such cloud-centric approaches, a large amount of data (e.g., images and videos) will be transferred between the end devices and the remote cloud datacenter backwards and forwards via a long wide-area network, which may potentially result in intolerable latency and extravagant energy.
To alleviate this problem, we exploit the emerging edge computing paradigm.
The principal idea of edge computing \cite{mao2017survey, shi2016edge, mach2017mobile} is to sink the cloud computing capability from the network core to the network edges (e.g., base stations and WLAN) in close proximity to end devices \cite{jararweh2017software, chen2018thriftyedge, yang2018meets, ouyang2018follow, chen2018efficient, xu2018distilling}.
This novel feature enables computation-intensive and latency-critical DNN-based applications to be executed in a real-time responsive manner (i.e., edge intelligence)\cite{zhou2019edge}.
By leveraging edge computing, we can design an on-demand low-latency DNNs inference framework for supporting real-time edge AI applications.

While recognizing the benefits of edge intelligence, our empirical study reveals that the performance of edge-based DNN inference is still highly sensitive to the available bandwidth between the edge server and the mobile device.
Specifically, as the bandwidth drops from 1Mbps to 50kbps, the edge-based DNN inference latency increases from 0.123s to 2.317s (detailed in Sec. \ref{sec:ineff}).
Considering the vulnerable and volatile network environment in practical deployment, a natural question is whether we can further optimize the DNN inference under the versatile network environment, especially for some mission-critical DNN-based applications such as intelligent security and industrial robotics \cite{qualcomm2017arvr}.

On this issue, in this paper, we exploit the edge computing paradigm and propose \textsf{Edgent}, a low-latency co-inference framework via device-edge synergy.
Towards low-latency edge intelligence\footnote{As an initial study, in this work, we focus on the problem of execution latency optimization. Energy efficiency will be considered in future studies.}, \textsf{Edgent} pursues two design knobs.
The first is DNN partitioning, which adaptively partitions DNN computation between mobile devices and the edge server according to the available bandwidth so as to utilize the computation capability of the edge server. However, it is insufficient to meet the stringent responsiveness requirement of some mission-critical applications since the execution performance is still restrained by the rest of the model running on mobile devices.
Therefore, \textsf{Edgent} further integrate the second knob, DNN right-sizing, which accelerates DNN inference by early exiting inference at an intermediate DNN layer.
Essentially, the early-exit mechanism involves a latency-accuracy tradeoff.
To strike a balance on the tradeoff with the existing resources, \textsf{Edgent} makes a joint optimization on both DNN partitioning and DNN right-sizing in an on-demand manner.
Specifically, for mission-critical applications that are typically with a pre-defined latency requirement, \textsf{Edgent} maximizes the accuracy while promising the latency requirement.

Considering the versatile network condition in practical deployment, \textsf{Edgent} further develops a tailored configuration mechanism so that \textsf{Edgent} can pursue better performance in both static network environment and dynamic network environment.
Specifically, in a static network environment (e.g., local area network with fiber or mmWave connection), we regard the bandwidth as stable and figure out a collaboration strategy through execution latency estimation based on the current bandwidth.
In this case, \textsf{Edgent} trains regression models to predict the layer-wise inference latency and accordingly derives the optimal configurations for DNN partitioning and DNN right-sizing.
In a dynamic network environment (e.g., 5G cellular network, vehicular network),
to alleviate the impact of network fluctuation, we build a look-up table by profiling and recording the optimal selection of each bandwidth state and specialize the runtime optimizer to detect the bandwidth state transition and map the optimal selection accordingly.
Through our specialized design for different network environment, \textsf{Edgent} is able to maximize the inference accuracy without violating the application responsiveness requirement.
The prototype implementation and extensive evaluations based on the Raspberry Pi and the desktop PC demonstrate \textsf{Edgent}'s effectiveness in enabling on-demand low-latency edge intelligence.

To summarize, we present the contribution of this paper as follows:
\begin{itemize}
	\item  We propose \textsf{Edgent}, a framework for on-demand DNN collaborative inference through device-edge synergy, in which we jointly optimize DNN partitioning and DNN right-sizing to maximize the inference accuracy while promising application latency requirement.
	\item  Considering the versatile network environments (i.e., static network environment and dynamic network environment), we specialize the workflow design for \textsf{Edgent} to achieve better performance.
	\item We implement and experiment \textsf{Edgent} prototype with a Raspberry Pi and a desktop PC. The evaluation results based on the real-world network trace dataset demonstrate the effectiveness of the proposed \textsf{Edgent} framework.
\end{itemize}

The rest of this paper is organized as follows. First, we review the related literature in Sec. \ref{related work} and present the background and motivation in Sec. \ref{bgm}.
Then we propose the design of \textsf{Edgent} in Sec. \ref{framework intro}.
The results of the performance evaluation are shown in Sec. \ref{exp res} to demonstrate the effectiveness of \textsf{Edgent}.
Finally, we conclude in Sec. \ref{conclu}.

\section{Related Work}\label{related work}
Discussions on the topic of mobile DNN computation have recently obtained growing attention.
By hosting artificial intelligence on mobile devices, mobile DNN computation deploys DNN models close to users in order to achieve more flexible execution as well as more secure interaction \cite{zhou2019edge}.
However, it is challenging to directly execute the computation-intensive DNNs on mobile devices due to the limited computation resource. On this issue, existing efforts dedicate to optimize the DNN computation on edge devices.

Towards low-latency and energy-efficient mobile DNN computation, there are mainly three ways in the literature: runtime management, model architecture optimization and hardware acceleration.
Runtime management is to offload computation from mobile devices to the cloud or the edge server, which utilizes external computation resource to obtain better performance.
Model architecture optimization attempts to develop novel DNN structure so as to achieve desirable accuracy with moderate computation \cite{Balasubramanian2017MobiRNN,Wu2015Quantized,SqueezeNet,Lane2015DeepEar,Howard2017MobileNets}.
For example, aiming at reducing resource consumption during DNN computation, DNN models are compressed by model pruning \cite{Han2015Deep,Kim2016Compression,LaneDXTK}.
Recent advance in such kind of optimization has turned to Network Architecture Search (NAS) \cite{cai2018efficient, liu2018progressive, pham2018efficient}.
Hardware acceleration generally embraces basic DNN computation operations in hardware level design \cite{kim2019mulayer, umuroglu2017finn, reagen2016minerva} while some works aim at optimizing the utilization of the existing hardware resources \cite{graves2016hybrid, bateni2018apnet, islam2019zygarde}.

As one of the runtime optimization methods, DNN partitioning technology segments specific DNN model into some successive parts and deploys each part on multiple participated devices.
Specifically, some frameworks \cite{Kang2017Neurosurgeon, Eshratifar2018JointDNN, li2018learning} take advantage of DNN partitioning to optimize the computation offloading between the mobile devices and the cloud, while some framework target to distribute computation workload among mobile devices \cite{Xu2017Enabling, Hadidi2018Musical,zhou2019aaiot}.
Regardless of how many devices are involved, DNN partitioning dedicates to maximize the utilization of external computation resources so as to accelerate the mobile computation.
As for DNN right-sizing, it focuses on adjusting the model size under the limitation of the existing environment.
On this objective, DNN right-sizing appeals to specialized training technique to generate the deuterogenic branchy DNN model from the original standard DNN model.
In this paper, we implement the branchy model with the assist of the open-source BranchyNet \cite{7900006} framework and Chainer \cite{chainer} framework.

Compared with the existing work, the novelty of our framework is summarized in the following three aspects.
First of all, given the pre-defined application latency requirement, \textsf{Edgent} maximizes the inference accuracy according to the available computation resources, which is significantly different from the existing studies. This feature is essential for practical deployment since different DNN-based applications may require different execution deadlines under different scenarios.
Secondly, \textsf{Edgent} integrates both DNN partitioning and DNN right-sizing to maximize the inference accuracy while promising application execution deadline.
It is worth noting that neither the model partitioning nor the model right-sizing can well address the timing requirement challenge solely. For model partitioning, it does reduce the execution latency, whereas the total processing time is still restricted by the part on the mobile device.
For model right-sizing, it accelerates the inference processing through the early-exit mechanism, but the total computation workload is still dominated by the original DNN model architecture and thus it is hard to finish the inference before the application deadline.

Therefore, we propose to integrate these two approaches to expand the design space.
The integration of model partitioning and model right-sizing is not a one-step effort, and we need to carefully design the decision optimization algorithms to fully explore the selection of the partition point and the exiting point and thus to strike a good balance between accuracy and latency in an on-demanded manner.
Through these efforts, we can achieve the design target such that given the predefined latency constraint, the DNN inference accuracy is maximized without violating the latency requirement.
Last but not least, we specialize the design of \textsf{Edgent} for both static and dynamic network environments while existing efforts (e.g., \cite{Hadidi2018Musical}) mainly focus on the scenario with stable network. Considering the diverse application scenarios and deployment environments in practice, we specialize the design of the configurator and the runtime optimizer for the static and dynamic network environments, by which \textsf{Edgent} can generate proper decisions on the exit point and partition point tailored to the network conditions.

\section{Background and Motivation} \label{bgm}
\label{motivation}
In this section, we first give a brief introduction on DNN. Then we analyze the limitation of edge- and device-only methods, motivated by which we explore the way to utilize DNN partitioning and right-sizing to accelerate DNN inference with device-edge synergy.

\subsection{A Brief Introduction on DNN}

\begin{figure}[t]
	\centering
	\includegraphics[width=0.9\linewidth]{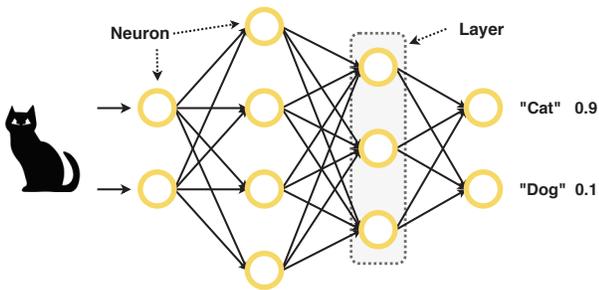}\\
	\caption{A four-layer DNN for image classification}\label{dnnexample}
\end{figure}

With the proliferation of data and computation capability, DNN has served as the core technology for a wide range of intelligent applications across Computer Vision (CV) \cite{szegedy2015going} and Natural Language Processing (NLP) \cite{Wang2015LSTM}. Fig. \ref{dnnexample} shows a toy DNN for image recognition that recognizes a cat. As we can see, a typical DNN model can be represented as a directed graph, which consists of a series of connected layers where neurons are connected with each other.
During the DNN computation (i.e., DNN training or DNN inference), each neuron accepts weighted inputs from its neighborhood and generates outputs after some activation operations.
A typical DNN may have dozens of layers and hundreds of nodes per layer and the total number of parameters can easily reach to the millions level, thus a typical DNN inference demands a large amount of computation.
In this paper, we focus on DNN inference rather than DNN training since the training process is generally delay-tolerant and often done offline on powerful cloud datacenters.

\subsection{Insufficiency of Device- or Edge-Only DNN Inference}
\label{sec:ineff}
\begin{figure}[t]
	\centering
	\includegraphics[width=0.5\linewidth]{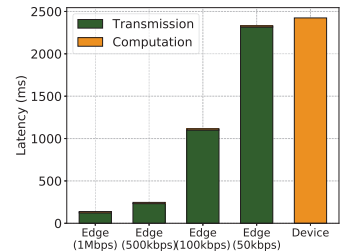}\\
	\caption{Execution runtime of edge-only (Edge) and device-only (Device) approaches of AlexNet under different bandwidths}\label{ineff}
\end{figure}

Traditional mobile DNN computation is either solely performed on mobile devices or wholly offloaded to cloud/edge servers. Unfortunately, both approaches may lead to poor performance (i.e., high end-to-end latency), making it difficult to meet real-time applications latency requirements \cite{chen2018thriftyedge}.
For illustration, we employ a Raspberry Pi and a desktop PC to emulate the mobile device and edge server respectively, and perform image recognition task over cifar-10 dataset with the classical AlexNet model\cite{NIPS2012_4824}.
Fig. \ref{ineff} depicts the end-to-end latency subdivision of different methods under different bandwidths on the edge server and the mobile device (simplified as Edge and Device in Fig. \ref{ineff}).
As shown in Fig. \ref{ineff}, it takes more than 2s to finish the inference task on a resource-constrained mobile device.
As a contrast, the edge server only spends 0.123s for inference under a 1Mbps bandwidth.
However, as the bandwidth drops, the execution latency of the edge-only method climbs rapidly (the latency climbs to 2.317s when the bandwidth drops to 50kbps).
This indicates that the performance of the edge-only method is dominated by the data transmission latency (the computation time at server side remains at $\sim$10ms) and is therefore highly sensitive to the available bandwidth.
Considering the scarcity of network bandwidth resources in practice (e.g., due to network resource contention between users and applications) and the limitation of computation resources on mobile devices, device- and edge-only methods are insufficient to support emerging mobile applications in stringent real-time requirements.

\subsection{DNN Partitioning and Right-Sizing towards Edge Intelligence}

\begin{figure}[t]
	\centering
	\includegraphics[width=\linewidth]{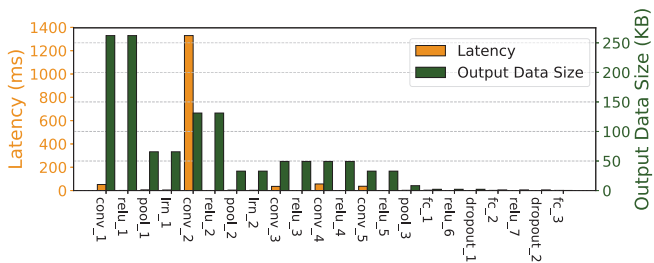}\\
	\caption{AlexNet layer-wise runtime and output data size on Raspberry Pi}\label{Runtime}
\end{figure}

\textbf{DNN Partitioning:} To better understand the performance bottleneck of DNN inference, in Fig. \ref{Runtime} we refine the layer-wise execution latency (on Raspberry Pi) and the intermediate output data size per layer.
Seen from Fig. \ref{Runtime}, the latency and output data size of each layer show great heterogeneity, implying that a layer with a higher latency may not output larger data size.
Based on this observation, an intuitive idea is DNN partitioning, i.e., dividing DNN into two parts and offloading the computation-intensive part to the server at a low transmission cost and thus reducing total end-to-end execution latency.
As an illustration, we select the second local response normalization layer (i.e., \texttt{lrn\_2}) in Fig. \ref{Runtime} as the partition point and the layers before the point are offloaded to the server side while the rest remains on the device.
Through model partitioning between device and edge, hybrid computation resources in proximity can be in comprehensive utilization towards low-latency DNN inference.

\begin{figure}[t]
	\centering
	\includegraphics[width=0.8\linewidth]{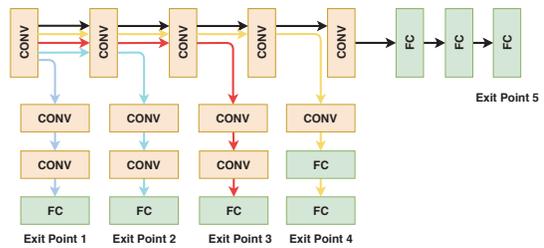}\\
	\caption{A branchy AlexNet model with five exit points}\label{Branchy AlexNet}
\end{figure}

\textbf{DNN Right-Sizing:} Although DNN partitioning can significantly reduce the latency, it should be noted that with the optimal DNN partitioning, the inference latency is still constrained by the remaining computation on the mobile device.
To further reduce the execution latency, the DNN right-sizing method is employed in conjunction with DNN partitioning.
DNN right-sizing accelerates DNN inference through the early-exit mechanism.
For example, by training DNN models with multiple exit points, a standard AlexNet model can be derived as a branchy AlexNet as Fig. \ref{Branchy AlexNet} shows, where a shorter branch (e.g., the branch ended with the exit point 1) implies a smaller model size and thus a shorter runtime.
Note that in Fig. \ref{Branchy AlexNet} only the convolutional layers (CONV) and the fully-connected layers (FC) are drawn for the ease of illustration.
This novel branchy structure demands novel training methods.
In this paper, we implement the branchy model training with the assist of the open-source BranchyNet \cite{7900006} framework.

\textbf{Problem Definition:} Clearly, DNN right-sizing leads to a latency-accuracy tradeoff, i.e., while the early-exit mechanism reduces the total inference latency, it hurts the inference accuracy.
Considering the fact that some latency-sensitive applications have strict deadlines but can tolerate moderate precision losses, we can strike a good balance between latency and accuracy in an on-demand manner.
In particular, given the pre-defined latency requirement, our framework should maximize accuracy within the deadline.
More precisely, the problem addressed in this paper can be summarized as how to make a joint optimization on DNN partitioning and DNN right-sizing in order to maximize the inference accuracy without violating the pre-defined latency requirement.

\section{Framework and Design} \label{framework intro}
In this section, we present the design of \textsf{Edgent}, which generates the optimal collaborative DNN inference plan that maximizes the accuracy while meeting the latency requirement in both the static and dynamic bandwidth environment.

\subsection{Framework Overview}
\begin{figure*}[t]
	\centering
	\includegraphics[width=0.8\linewidth]{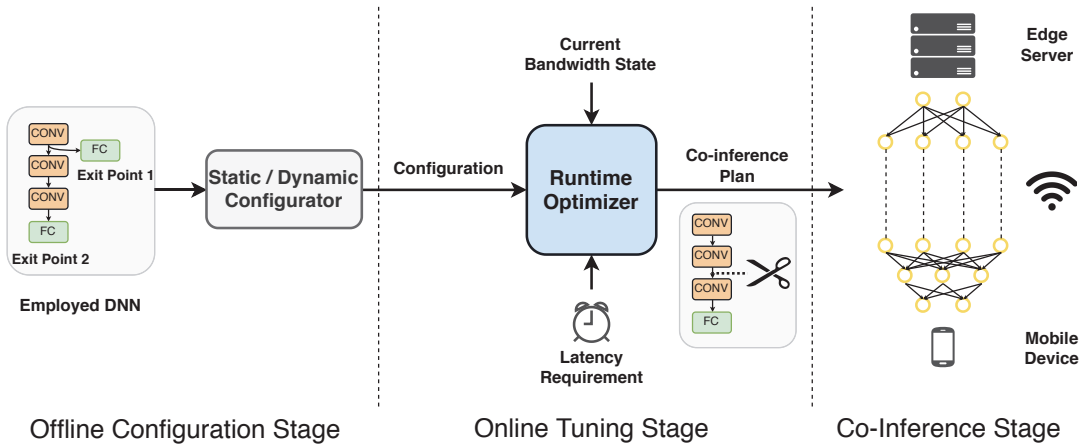}\\
	\caption{Edgent framework overview}\label{Edgent Overview}
\end{figure*}

\textsf{Edgent} aims to pursue a better DNN inference performance across a wide range of network conditions. As shown in Fig. \ref{Edgent Overview}, \textsf{Edgent} works in three stages: offline configuration stage, online tuning stage and co-inference stage.

At the \textbf{offline configuration stage}, \textsf{Edgent} inputs the employed DNN to the \emph{Static/Dynamic Configurator} component and obtains the corresponding configuration for online tuning.
To be specific, composed of trained regression models and branchy DNN model, the static configuration will be employed where the bandwidth keeps stable during the DNN inference (which will be detailed in Sec. \ref{edgent static}), while composed of the trained branchy DNN and the optimal selection for different bandwidth states, the dynamic configuration will be used adaptive to the state dynamics (which will be detailed in Sec. \ref{edgent dynamic}).

At the \textbf{online tuning stage}, \textsf{Edgent} measures the current bandwidth state and makes a joint optimization on DNN partitioning and DNN right-sizing based on the given latency requirement and the configuration obtained offline, aiming at maximizing the inference accuracy under the given latency requirement.

At the \textbf{co-inference stage}, based on the co-inference plan (i.e., the selected exit point and partition point) generated at the online tuning stage, the layers before the partition point will be executed on the edge server with the rest remaining on the device.

During DNN inference, the bandwidth between the mobile device and the edge server may be relatively stable or frequently changing. Though \textsf{Edgent} runs on the same workflow in both static and dynamic network environments, the function of \emph{Configurator} component and \emph{Runtime Optimizer} component differ.
Specifically, under a static bandwidth environment, the configurator trains regression models to predict inference latency and the branchy DNN to enable early-exit mechanism.
The static configuration generated offline includes the trained regression models and the trained branchy DNN, based on which the \emph{Runtime Optimizer} will figure out the optimal co-inference plan.
Under a dynamic bandwidth environment, the dynamic configurator creates a configuration map that records the optimal selection for different bandwidth state via the change point detector, which will then be input to the \emph{Runtime Optimizer} to generate the optimal co-inference plan.
In the following, we will discuss the specialized design of the configurator and optimizer for the static and dynamic environments, respectively.

\subsection{\textsf{Edgent} for Static Environment}\label{edgent static}

\begin{figure}[t]
	\centering
	\includegraphics[width=\linewidth]{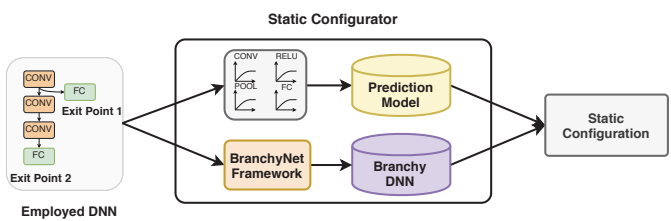}\\
	\caption{The static configurator of \textsf{Edgent}}\label{static conf}
\end{figure}

As a starting point, we first consider our framework design in the case of the static network environment.
The key idea of the static configurator is to train regression models to predict the layer-wise inference latency and train the branchy model to enable early-exit mechanism.
The configurator specialized for the static bandwidth environment is shown in Fig. \ref{static conf}.

At the \textbf{offline configuration stage}, to generate static configuration, the static configurator initiates two tasks: (1) profile layer-wise inference latency on the mobile device and the edge server respectively and accordingly train regression models for different kind of DNN layer (e.g., \texttt{Convolution}, \texttt{Fully-Connected}, etc.), (2) train the DNN model with multiple exit points via BranchyNet framework to obtain branchy DNN.
The profiling process is to record the inference latency of each type of layers rather than that of the entire model.
Base on the profiling results, we establish the prediction model for each type of layers by performing a series of regression models with independent variables shown in Table. \ref{independent var}.
Since there are limited types of layers in a typical DNN, the profiling cost is moderate.
Since the layer-wise inference latency is infrastructure-dependent and the DNN training is application-related, for a specific DNN inference \textsf{Edgent} only needs to initialize the above two tasks for once.

\begin{table}[t]
	\centering
	\caption{The independent variables of prediction models}
	\label{independent var}
	\footnotesize
    \begin{tabular}{c|c}
		\hline
		\textbf{Type of DNN Layer} & \textbf{Independent Variable(s)} \\ \hline
		Convolutional & \begin{tabular}[c]{@{}l@{}}\# of input feature maps,\\ (filter size/stride)$^2\times$(\# of filters)\end{tabular}\\ \hline
		Relu & input data size \\ \hline
		Pooling & \begin{tabular}[c]{@{}l@{}}input data size, output data size\end{tabular} \\ \hline
		Local Response Normalization & input data size \\ \hline
		Dropout & input data size \\ \hline
		Fully-Connected & \begin{tabular}[c]{@{}l@{}}input data size, output data size\end{tabular} \\ \hline
	\end{tabular}
\end{table}

At the \textbf{online tuning stage}, using the static configuration (i.e., the prediction model and the branchy DNN), the \emph{Runtime Optimizer} component searches the optimal exit point and partition point to maximize the accuracy while ensuring the execution deadline with three inputs: (1) the static configuration, (2) the measured bandwidth between the edge server and the end device and (3) the latency requirement.
The search process of joint optimization on the selection of partition point and exit point is described in \emph{Algorithm 1}.
For a DNN model with $M$ exit points, we denote that the branch of $i\text{-}th$ exit point has $N_i$ layers and $D_{p}$ is the output of the $p\text{-}th$ layer.
We use the abvoe-mentioned regression models to predict the latency $ED_{j} $ of the $j\text{-}th$ layer running on the device and the latency $ES_{j}$ of the $j\text{-}th$ layer running on the server.
Under a certain bandwidth $B$, fed with the input data $Input$, we can calculate the total latency $A_{i, p}$ by summing up the computation latency on each side and the communication latency for transferring input data and intermediate execution result.
We denote $p\text{-}th$ layer as the partition point of the branch with the $i\text{-}th$ exit point.
Therefore, $p=1$ indicates that the total inference process will only be executed on the device side (i.e., $ES_{p}=0, D_{p-1}/B=0, Input/B=0$) whereas $p=N_{i}$ means the total computation is only done on the server side (i.e., $ED_{p}=0, D_{p-1}/B=0$).
Through the exhaustive search on points, we can figure out the optimal partition point with the minimum latency of the $i\text{-}th$ exit point.
Since the model partitioning does not affect the inference accuracy, we can successively test the DNN inference with different exit layers (i.e., with different precision) and find the model with the maximum accuracy while satisfying the latency requirement at the same time.
As the regression models for layer-wise latency prediction have been trained in advance, \emph{Algorithm 1} mainly involves linear search operations and can be completed very quickly (no more than 1ms in our experiment).

There are two basic assumptions for our design. One is that we assume the existing DNN inference on mobile devices cannot satisfy the application latency requirement and there is an edge server in proximity that is available to be employed to accelerate DNN inference through computation offloading. The other assumption is that the regression models for performance prediction are trained based on the situation that the computation resources for the DNN model execution on the mobile device and the edge server are fixed and allocated beforehand. Nevertheless, these assumptions can be further relaxed, since we can train more advanced performance prediction models (e.g., using deep learning models) by taking different resource levels into account.

\begin{algorithm}[t]
	\scriptsize
	\caption{Runtime Optimizer for Static  Environment}
	\begin{flushleft}
		\hspace*{0.02in} {\bf Input:}
		\\ $M$: the number of exit points in the DNN model
		\\ $\{N_{i}|i=1,\cdots ,M \}$: the number of layers in the branch of exit pint $i$
		\\ $\{L_{j}|j=1,\cdots ,N_{i}\}$: the layers in the branch of exit point $i$
		\\ $\{D_{j}|j=1,\cdots ,N_{i}\}$: layer-wise output data size in the branch of exit point $i$
		\\ $f(L_{j})$: the prediction model that returns the $j\text{-}th$ layer's latency
		\\ $B$: current available bandwidth
		\\ $Input$: input data size
		\\ $Latency$: latency requirement\\
		\hspace*{0.02in} {\bf Output:}
		\\ Selection of exit point and partition point
	\end{flushleft}
	
	\begin{algorithmic}[1]
		\State \textbf{Procedure}
		\For{$i=M,\cdots ,1$}
		\State Select the branch of $i\text{-}th$ exit point
		\For{$j = 1,\cdots ,N_{i}$}
		\State $ES_{j}\gets f_{edge}(L_{j})$
		\State $ED_{j}\gets f_{device}({L_{j}})$
		\EndFor
		\State $A_{i, p} = \argmin\limits_{p=1,\cdots ,N_{i}} (\sum_{j=1}^{p-1}ES_{j} + \sum_{k=p}^{N_{i}}ED_{j} + Input/B + D_{p-1}/B)$
		\If{$A_{i, p} \leq Latency$}
		\State \Return Exit point $i$ and partition point $p$
		\EndIf
		\EndFor
		\State \Return NULL \Comment can not meet the latency requirement
	\end{algorithmic}
\end{algorithm}

\subsection{\textsf{Edgent} for Dynamic Environment}\label{edgent dynamic}
\begin{figure}[t]
	\centering
	\includegraphics[width=0.9\linewidth]{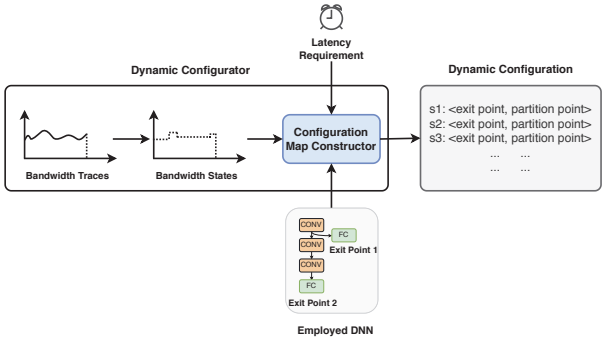}\\
	\caption{The dynamic configurator of \textsf{Edgent}}\label{dynamic conf}
\end{figure}

The key idea of \textsf{Edgent} for the dynamic environment is to exploit the historical bandwidth traces and employ \emph{Configuration Map Constructor} to generate the optimal co-inference plans for versatile bandwidth states in advance.
Specifically, under the dynamic environment \textsf{Edgnet} generates the dynamic configuration (i.e., a configuration map that records the optimal selections for different bandwidth states) at the offline stage, and at the online stage \textsf{Edgent} searches for the optimal partition plan according to the configuration map.
The configurator specialized for the dynamic bandwidth environment is shown in Fig. \ref{dynamic conf}.

At the \textbf{offline configuration stage}, the dynamic configurator performs following initialization: (1) sketch the bandwidth states (noted as $s1, s2, \cdots$) from the historical bandwidth traces and (2) pass the bandwidth states, latency requirement and the employed DNN to the static \textsf{Edgent} to acquire the optimal exit point and partition point for the current input.
The representation of the bandwidth states is motivated by the existing study of adaptive video streaming \cite{oboe2018}, where the throughput of TCP connection can be modeled as a piece-wise stationary process where the connection consists of multiple non-overlapping stationary segments.
In the dynamic configurator, it defines a bandwidth state $s$ as the mean of the throughput on the client side in a segment of the underlying TCP connection.
With each bandwidth state, we acquire the optimal co-inference plan by calling the \emph{Configuration Map Constructor} and record that in a map as the dynamic configuration.

The configuration map construction algorithm run in \emph{Configuration Map Constructor} is presented in \emph{Algorithm 2}.
The key idea of \emph{Algorithm 2} is to utilize the reward function to evaluate the selection of the exit point and partition point.
Since our design goal is to maximize the inference accuracy while promising the application latency requirement, it is necessary to measure whether the searched co-inference strategy can meet the latency requirement and whether the inference accuracy has been maximized.
Therefore, we define a reward to evaluate the performance of each search step as follow:
\begin{equation}
\label{eq:reward}
reward_{step}=\left\{
\begin{array}{rcl}
\exp{(acc)} + throughput, & & {t_{step} \leq t_{req}},\\
0, & & {else},
\end{array} \right.
\end{equation}
where $t_{step}$ is the average execution latency in the current search step (i.e., the selected exit point and partition point in the current search step), which equals to $\frac{1}{throughput}$.
The conditions of Equation (\ref{eq:reward}) prioritizes that the latency requirement $t_{req}$ should be satisfied, otherwise the reward will be set as $0$ directly. Whenever the latency requirement is met, the reward of the current step will be calculated as $\exp{(acc)} + throughput$, where $acc$ is the accuracy of current inference.
If the latency requirement is satisfied, the search emphasizes on improving the accuracy and when multiple options have the similar accuracy, the one with the higher throughput will be selected.
In \emph{Algorithm 2}, $s_{i}$ represents a bandwidth state extracted from the bandwidth traces and $C_{j}$ is a co-inference strategy (i.e., a combination of exit point and partition point ) indexed by $j$. $R(C_{j})$ denotes the reward of the co-inference strategy $C_j$, which can be obtained by calculating Equation (\ref{eq:reward}) according to the accuracy and the throughput of $C_j$.

\begin{algorithm}[htp]
	\scriptsize
	\caption{Configuration Map Construction}
	\begin{flushleft}
		\hspace*{0.02in} {\bf Input:}
		\\ $\{s_{i}|i=1,\cdots ,N \}$: the bandwidth states
		\\ $\{C_{j}|j=1,\cdots ,M \}$: the co-inference strategy
		\\ $R(C_{j})$: the reward of co-inference strategy $C_{j}$ \\
		
		\hspace*{0.02in} {\bf Output:}
		\\ Configuration Map
	\end{flushleft}
	
	\begin{algorithmic}[1]
		\State \textbf{Procedure}
		\For{$i=1,\cdots ,N$}
		\State Select the bandwidth state $s_{i}$
		
		\State $reward_{max} = 0$, $C_{optimal} = 0$
		\For{$j = 1,\cdots ,M$}
		\State $reward_{c_j}\gets R(C_{j})$
		\If{$reward_{max} \leq reward_{c_j}$}
		\State $reward_{max} = reward_{c_j}$, $C_{optimal} = C_{j}$
		\EndIf
		\EndFor
		
		\State Get the corresponding $exit \, point$ and $partition \, point$ of $C_{optimal}$
		\State Add $S_{i} : <exit \, point, \, partition \, point>$ to the Configuration Map
		\EndFor
		\State \Return Configuration Map
	\end{algorithmic}
\end{algorithm}

At the \textbf{online tuning stage}, the \emph{Runtime Optimizer} component selects the optimal co-inference plan according to the dynamic configuration and real-time bandwidth measurements.
\emph{Algorithm 3} depicts the whole process in \emph{Runtime Optimizer}.
Note that \emph{Algorithm 3} calls the change point detection function $D(B_{1,\cdots,t})$ \cite{adams2007bayesian} to detect the distributional state change of the underlying bandwidth dynamics.
Particularly, when the sampling distribution of the bandwidth measurement has changed significantly, the change point detection function records a change point and logs a bandwidth state transition.
Then with $find(state)$ function, the \emph{Runtime Optimizer} captures the corresponding co-inference strategy to the current bandwidth state (or the closest state) in the dynamic configuration and accordingly guides the collaborative inference process at the co-inference stage.

\section{Performance Evaluation} \label{exp res}
In this section, we present our implementation of \textsf{Edgent} and the evaluation results.

\subsection{Experimental Setup}
We implement a prototype based on the Raspberry Pi and the desktop PC to demonstrate the feasibility and efficiency of \textsf{Edgent}.
Equipped with a quad-core 3.40 GHz Intel processor and 8 GB RAM, a desktop PC is served as the edge server. Equipped with a quad-core 1.2 GHz ARM processor and 1 GB RAM, a Raspberry Pi 3 is used to act as a mobile device.

To set up a static bandwidth environment, we use the WonderShaper tool \cite{wondershaper} to control the available bandwidth. As for the dynamic bandwidth environment setting, we use the dataset of Belgium 4G/LTE bandwidth logs \cite{vanderHooft2016} to emulate the online dynamic bandwidth environment. To generate the configuration map, we use the synthetic bandwidth traces provided by Oboe \cite{oboe2018} to generate 428 bandwidth states range from 0Mbps to 6Mbps.

To obtain the branchy DNN, we employ BranchyNet \cite{7900006} framework and Chainer \cite{chainer} framework, which can well support multi-branchy DNN training.
In our experiments we take the standard AlexNet \cite{NIPS2012_4824} as the toy model and train the AlexNet model with five exit points for image classification over the cifar-10 dataset \cite{Krizhevsky2009Learning}. As shown in Fig. \ref{Branchy AlexNet}, the trained branchy AlexNet has five exit points, with each point corresponds to a branch of the standard AlexNet.
From the longest branch to the shortest branch, the number of layers in each exit point is 22, 20, 19, 16 and 12, respectively.

\begin{algorithm}[t]
	\scriptsize
	\caption{Runtime Optimizer for Dynamic Environment}
	\begin{flushleft}
		\hspace*{0.02in} {\bf Input:}
		\\ $\{B_{1,\cdots,t}\}$: the accumulated bandwidth measurements until the current moment $t$
	    \\ $\{C_{j}|j=1,\cdots, t \}$: the co-inference strategy
	    \\ $\{s_{i}|i=1,\cdots, t \}$: the bandwidth states
		\\ $D(B_{1,\cdots,t})$: the bandwidth state detection function that returns the current bandwidth state
		\\ $find(s)$: find the co-inference strategy corresponds to the given state $s$  \\

		\hspace*{0.02in} {\bf Output:}
		\\ Co-inference strategy
	\end{flushleft}
	
	\begin{algorithmic}[1]
		\State \textbf{Procedure}
        \State $C_{t} = C_{t-1}$
		\State $s_t = D(B_{i,\cdots,t})$
		\If{$s_{t} \neq s_{t-1}$}
		\State $ C_t \gets find(s_{t})$
		\EndIf
		\State $s_{t-1} = s_{t}$
		\State $C_{t-1} = C_{t}$
		\State \Return $C_t$
	\end{algorithmic}
\end{algorithm}

\subsection{Experiments in Static Bandwidth Environment} \label{static res}

In the static configurator, the prediction model for layer-wise prediction is trained based on the independent variables presented in Table. \ref{independent var}.
The branchy AlexNet is deployed on both the edge server and the mobile device for performance evaluation. Specifically, due to the high-impact characteristics of the latency requirement and the available bandwidth during the optimization procedure, the performance of \textsf{Edgent} is measured under different pre-defined latency requirements and varying available bandwidth settings.

\begin{figure*}[!ht]
	\centering
	
	\subfigure[Selection under different bandwidths]{
		\includegraphics[width=0.28\linewidth]{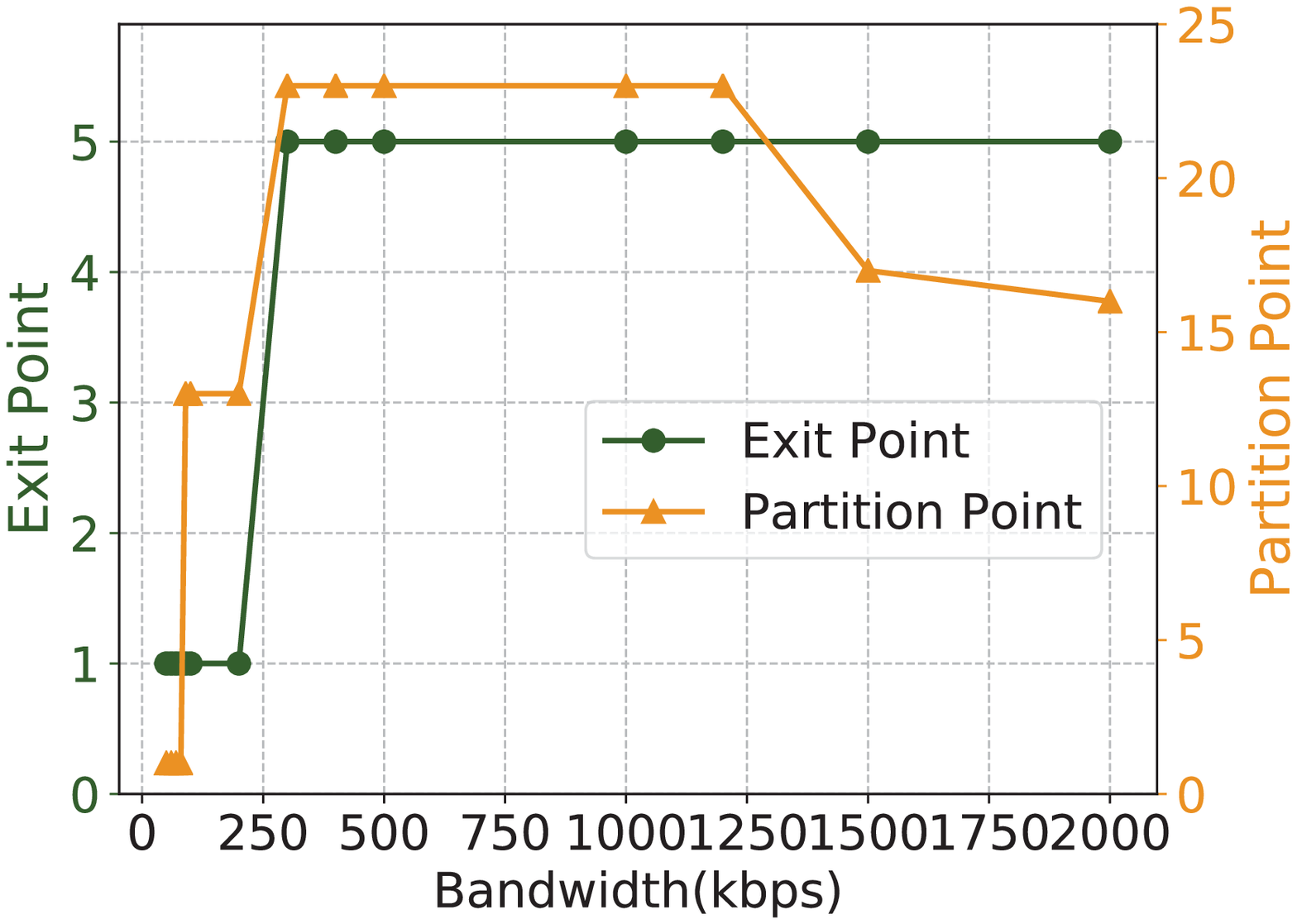}\label{exit and partition}
	}
	\subfigure[Model runtime under different bandwidths]{
		\includegraphics[width=0.3\linewidth]{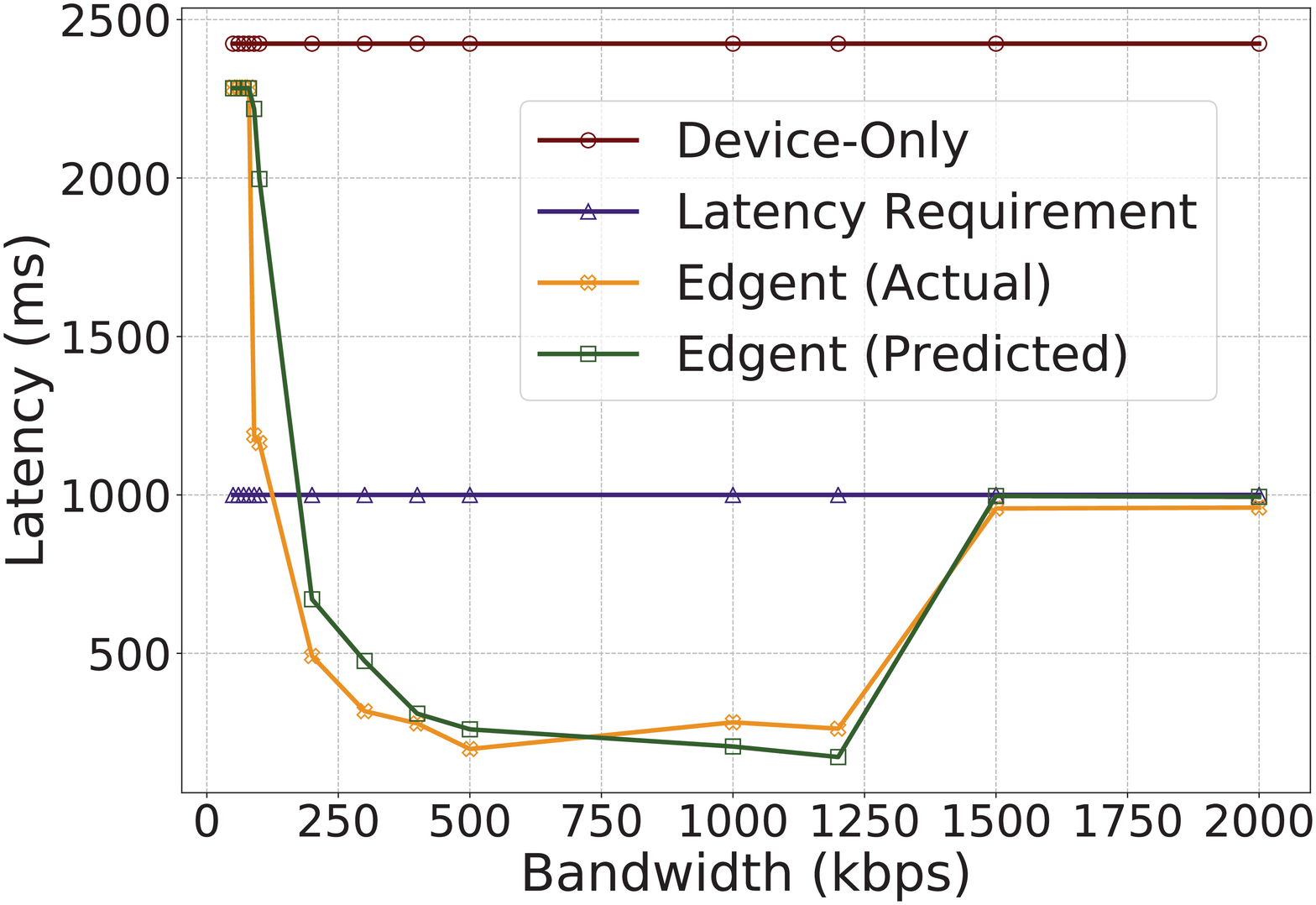}\label{model runtime}
	}
	\subfigure[Selection under different latency requirements]{
		\includegraphics[width=0.3\linewidth]{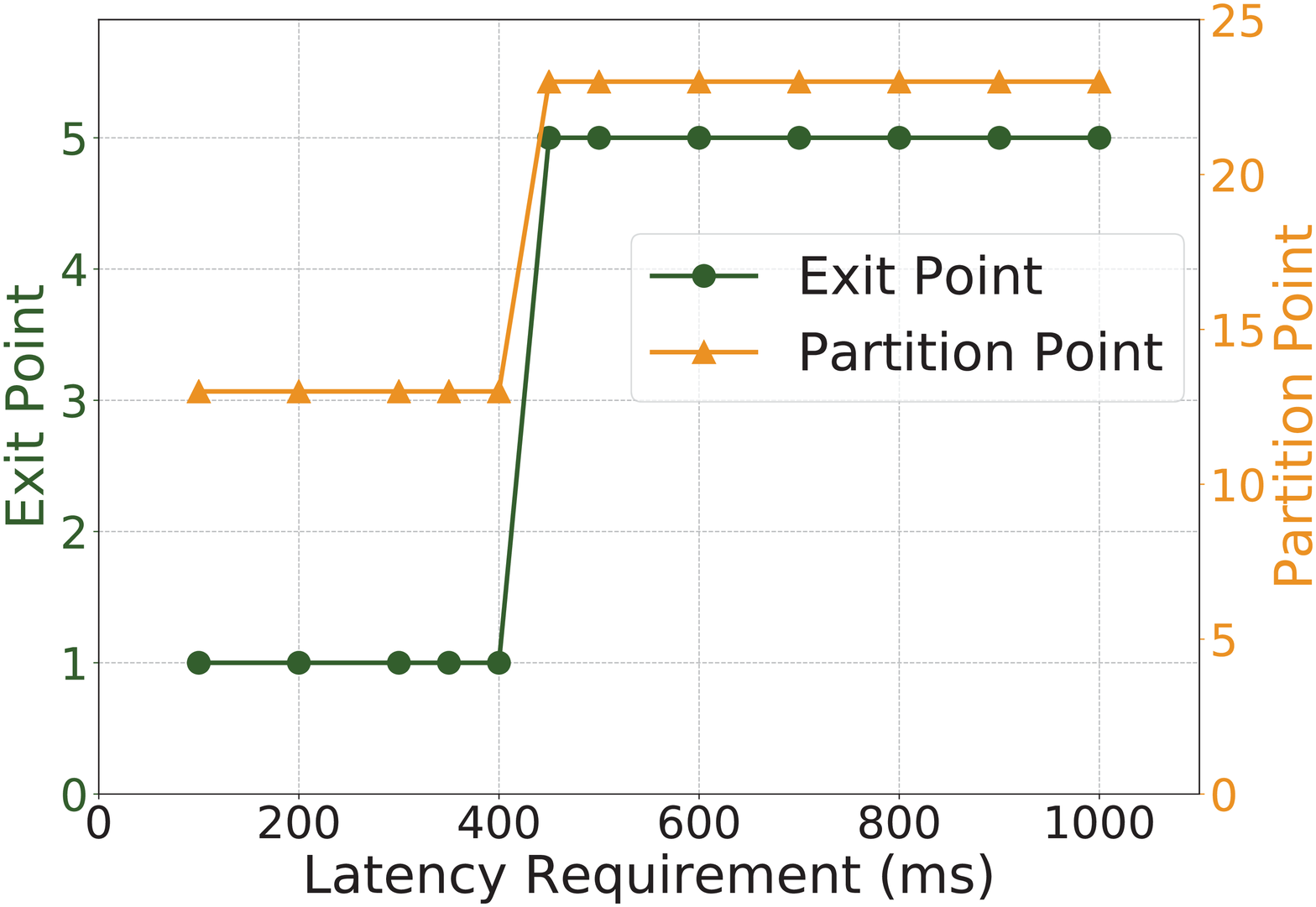}\label{exit and partition diff latency}
	}
	\caption{The results under different bandwidths and different latency requirements}\label{selections}
\end{figure*}

We first explore the impact of the bandwidth by fixing the latency requirement at 1000ms and setting the bandwidth from 50kbps to 1.5Mbps.
Fig. \ref{exit and partition} shows the optimal co-inference plan (i.e., the selection of partition point and exit point) generated by \textsf{Edgent} under various bandwidth settings.
Shown in Fig. \ref{exit and partition}, as bandwidth increases, the optimal exit point becomes larger, indicating that a better network environment leads to a longer branch of the employed DNN and thus higher accuracy.
Fig. \ref{model runtime} shows the inference latency change trend where the latency first descends sharply and then climbs abruptly as the bandwidth increases.
This fluctuation makes sense since the bottleneck of the system changes as the bandwidth becomes higher.
When the bandwidth is smaller than 250kbps, the optimization of \textsf{Edgent} is restricted by the poor communication condition and prefers to trade the high inference accuracy for low execution latency, for which the exit point is set as 3 rather than 5.
As the bandwidth rises, the execution latency is no longer the bottleneck so that the exit point climbs to 5, implying a model with larger size and thus higher accuracy.
There is another interesting result that the curve of predicted latency and the measured latency is nearly overlapping, which shows the effectiveness of our regression-based prediction.
Next, we set the available bandwidth at 500kbps and vary the latency requirement from 1000ms to 100ms for further exploration.
Fig. \ref{exit and partition diff latency} shows the optimal partition point and exit point under different latency requirements.
As illustrated in Fig. \ref{exit and partition diff latency}, both the optimal exit point and partition point climb higher as the latency requirement relaxes, which means that a later execution deadline will provide more room for accuracy improvement.

\begin{figure}[t]
	\centering
	\includegraphics[width=\linewidth]{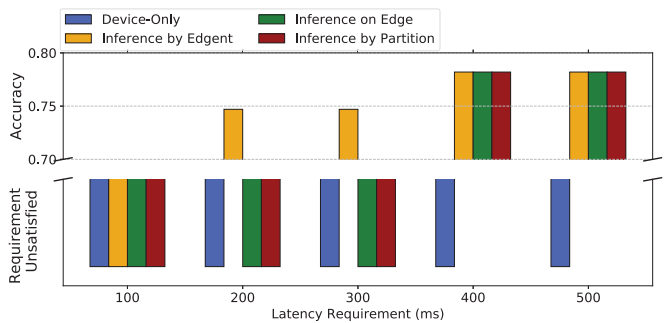}\\
	\caption{The accuracy comparison under various latency requirement}\label{accuracy comparison}
\end{figure}

Fig. \ref{accuracy comparison} shows the model inference accuracy of different methods under different latency requirement settings (the bandwidth is fixed as 400kbps).
The accuracy is described in negative when the method cannot satisfy the latency requirement.
As Fig. \ref{accuracy comparison} shows, given a tightly restrict latency requirement (e.g., 100ms), all the four methods fail to meet the requirement, for which all the four squares lay below the standard line.
However, as the latency requirement relaxes, \textsf{Edgent} works earlier than the other three methods (at the requirements of 200ms and 300ms) with the moderate loss of accuracy.
When the latency requirement is set longer than 400ms, all the methods except for device-only inference successfully finish execution in time.

\subsection{Experiments in Dynamic Bandwidth Environment} \label{dynamic res}

For the configuration map generation, we use the bandwidth traces provided in Oboe \cite{oboe2018}. Each bandwidth trace in the dataset consists of 49 pairs of data tuple about download chunks, including start time, end time and the average bandwidth.
We calculate the mean value of all the average bandwidth in the same bandwidth trace to represent the bandwidth state fluctuation, from which we obtain 428 bandwidth states range from 0Mbps to 6Mbps.
According to the \emph{Algorithm 2}, through the exhaustive search, we figure out the optimal selection of each bandwidth state.
The latency requirement in this experiment is also set to 1000ms.

For online change point detection, we use the existing implementation \cite{obcd} and integrate it with the \emph{Runtime Optimizer}.
We use the Belgium 4G/LTE bandwidth logs dataset \cite{vanderHooft2016} to perform online bandwidth measurement, which records the bandwidth traces that are measured on several types of transportation: on foot, bicycle, bus, train or car.
Additionally, Since that most of the bandwidth logs are over 6Mbps and in some cases even up to 95Mbps, to adjust the edge computing scenario, in our experiment, we scale down the bandwidth of the logs and limit it in a range from 0Mbps to 10Mbps.

\begin{figure}[htbp]
	\centering
	
	\subfigure[A example bandwidth trace on Belgium 4G/LTE dataset \cite{vanderHooft2016}]{
		\includegraphics[width=0.8\linewidth]{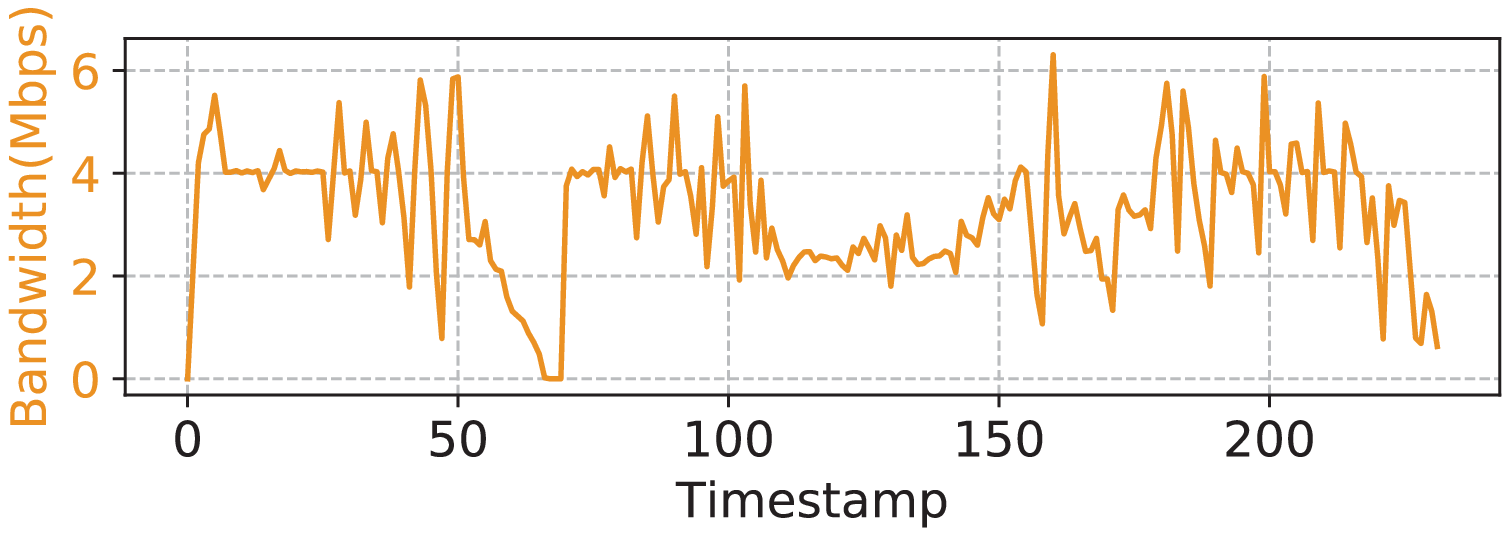}\label{bus trace}
	}
	\subfigure[The throughput of DNN model inference]{
		\includegraphics[width=0.8\linewidth]{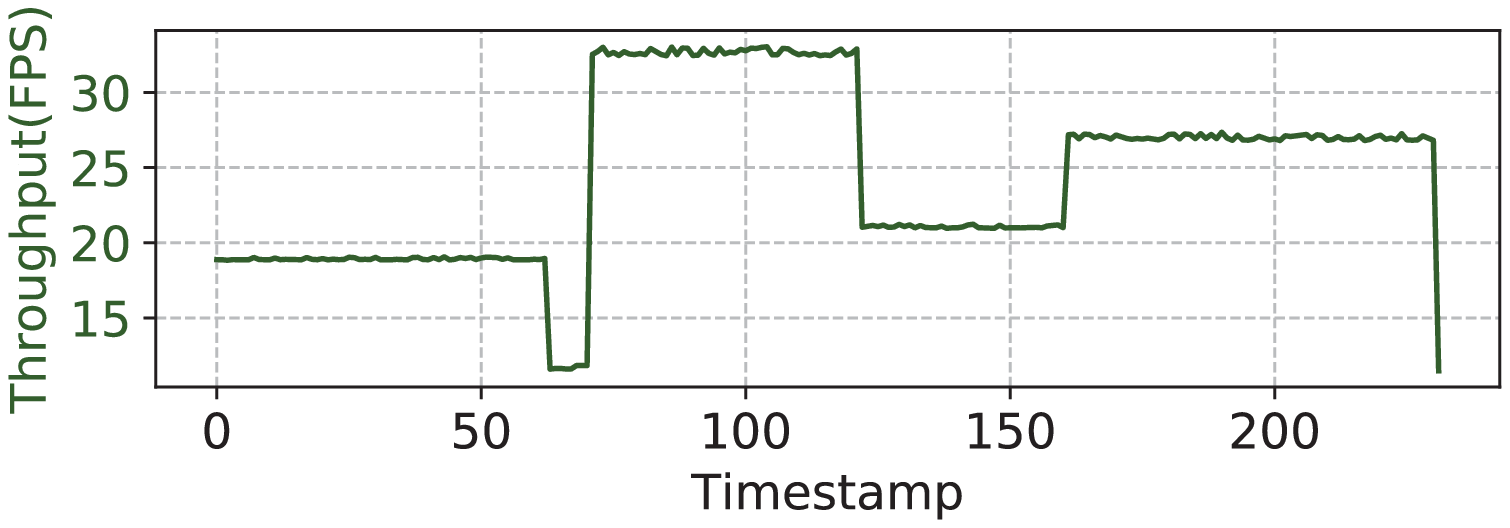}\label{bus throughput}
	}
	\subfigure[The selection of exit point and partition point at each timestamp in the bandwidth traces]{
		\includegraphics[width=0.8\linewidth]{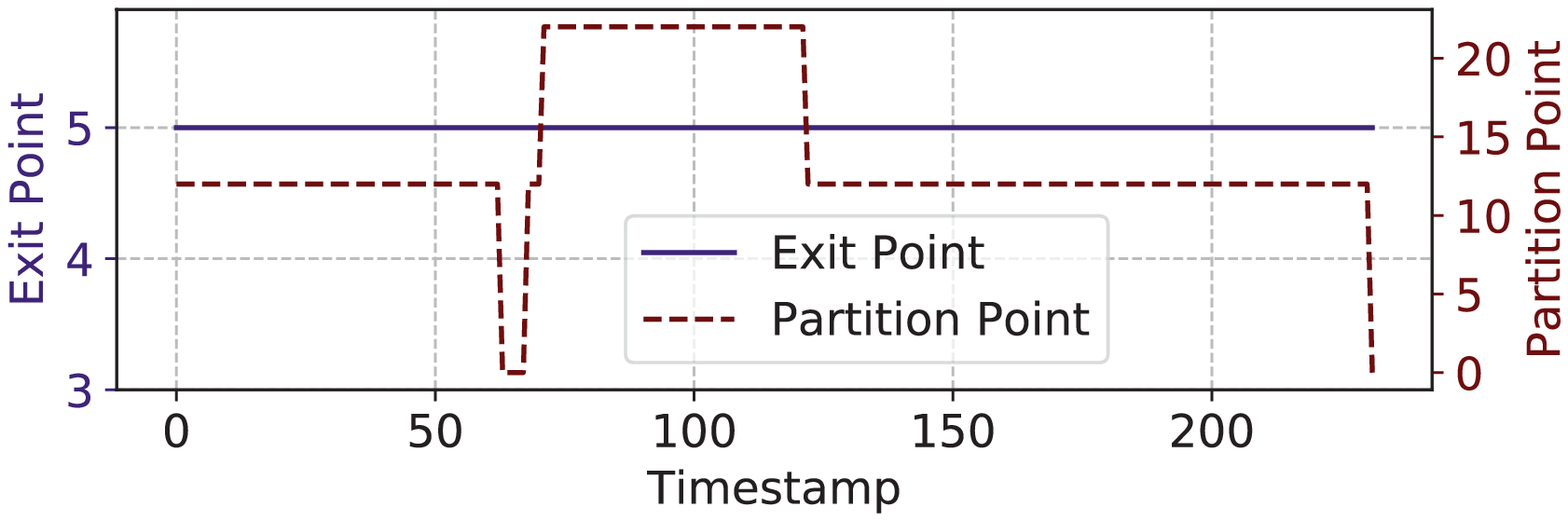}\label{bus selection}
	}
	\caption{An example showing the decision-making ability of \textsf{Edgent} in a bandwidth traces recorded on bus}\label{bus result}
\end{figure}

In this experiment \textsf{Edgent} runs in a dynamic bandwidth environment emulated by the adjusted Belgium 4G/LTE bandwidth logs. Fig. \ref{bus trace} shows an example bandwidth trace on the dataset that is recorded on a running bus.
Fig. \ref{bus throughput} shows the DNN model inference throughput results under the bandwidth environment showed in Fig. \ref{bus trace}.
The corresponding optimal selection of the exit point and partition point is presented in Fig. \ref{bus selection}.
Seen from Fig. \ref{bus selection}, the optimal selection of model inference strategy varies with the bandwidth changes but the selected exit point stays at 5, which means that the network environment is good enough for \textsf{Edgent} to satisfy the latency requirement though the bandwidth fluctuates.
In addition, since the exit point remains invariable, the inference accuracy also keeps stable.
Dominated by our reward function design, the selection of partition points approximately follows the traces of the throughput result.
The experimental results show the effectiveness of \textsf{Edgent} under the dynamic bandwidth environment.

\begin{figure}[t]
	\centering
	\subfigure[Throughput]{
		\includegraphics[width=0.46\linewidth]{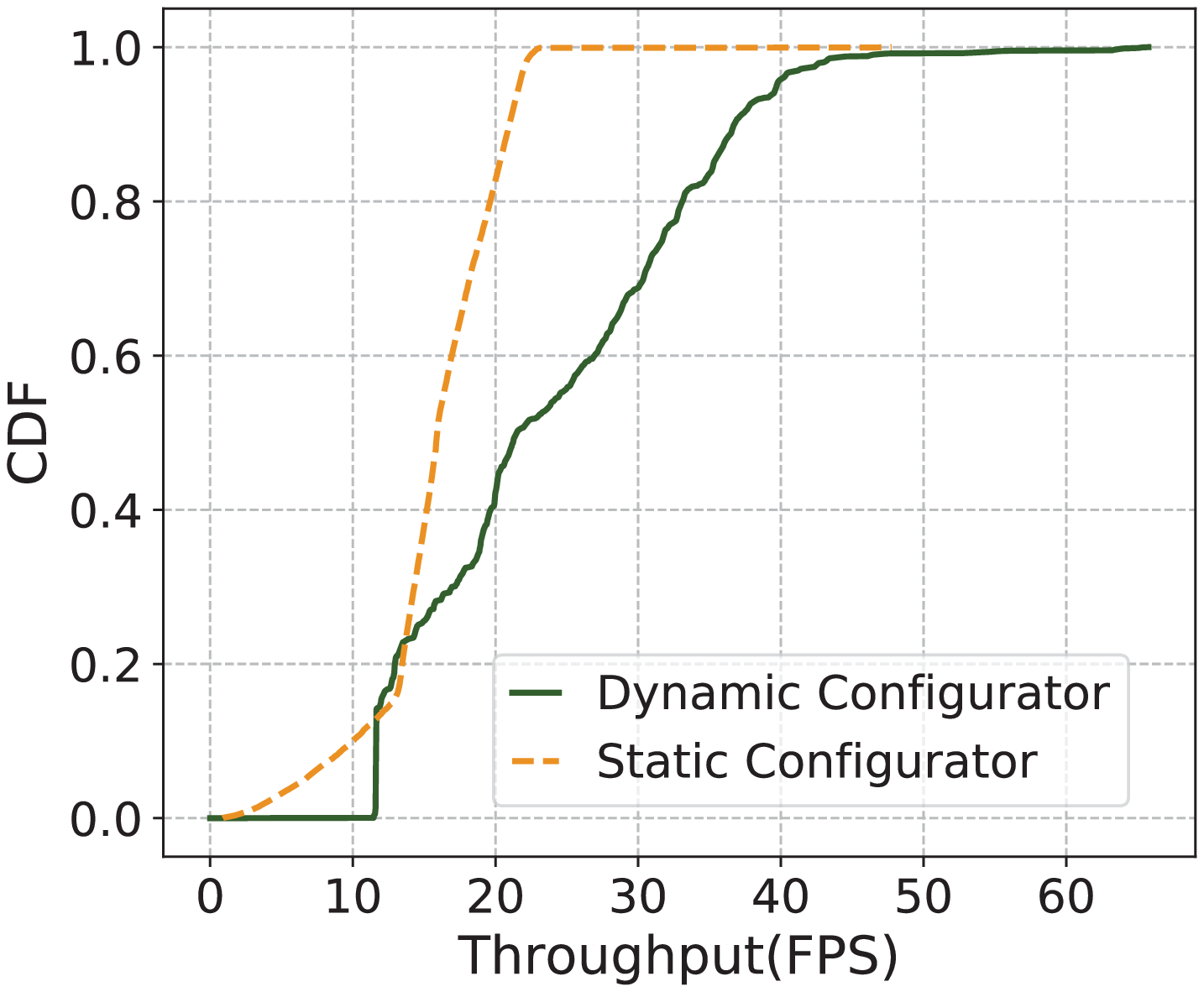}\label{conf comparison}
	}
	\subfigure[Reward]{
		\includegraphics[width=0.46\linewidth]{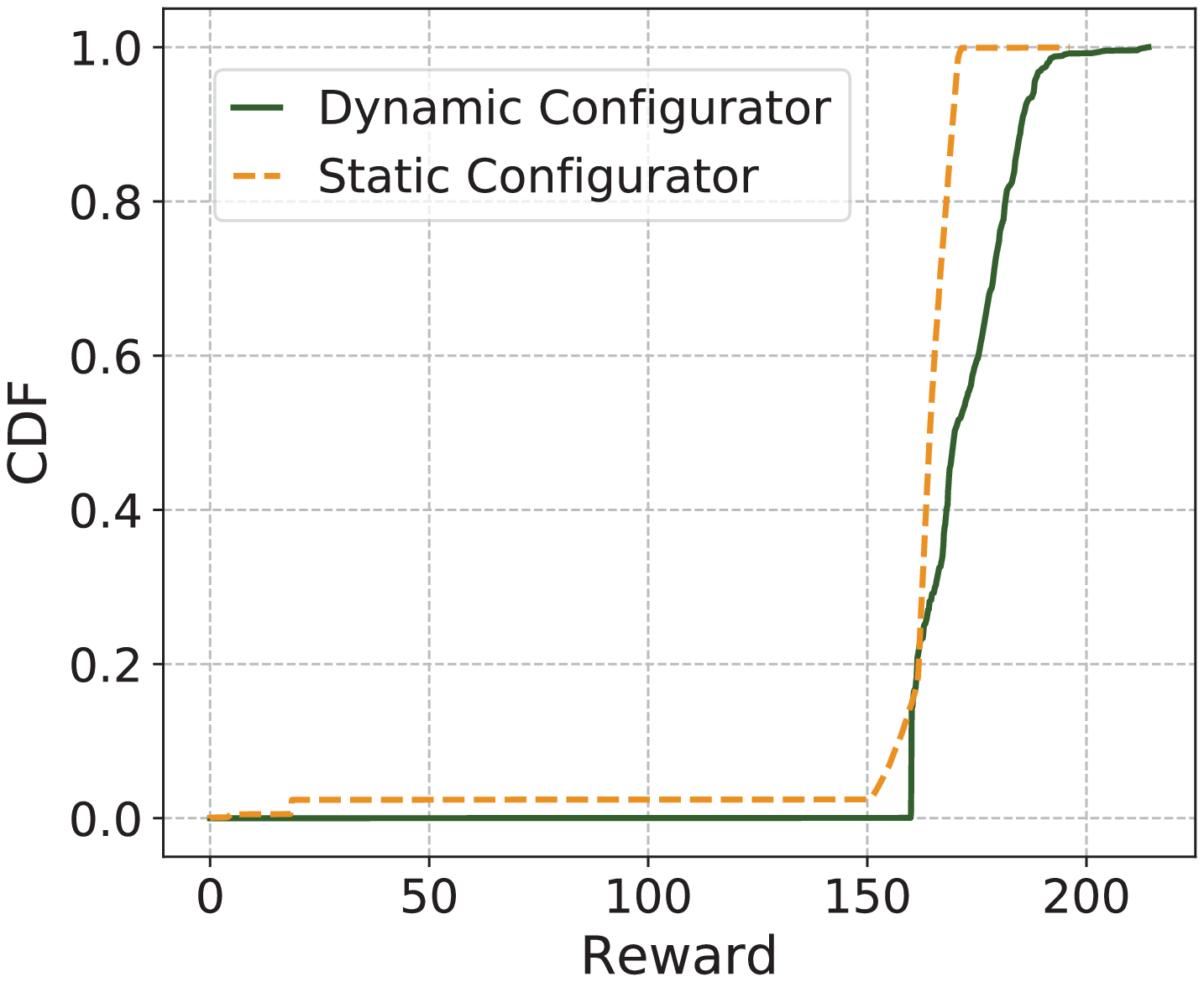}\label{reward comparison}
	}
	\caption{The throughput and reward comparison between two configurations}\label{tp conf}
\end{figure}

We further compare the static configurator and the dynamic configurator under the dynamic bandwidth environment in Fig. \ref{tp conf}.
We set the latency requirement as 1000ms and record the throughput and the reward for the two configurators, base on which we calculate the Cumulative Distribution Function (CDF).
Seen from Fig. \ref{conf comparison}, under the same CDF level, \textsf{Edgent} with the dynamic configurator achieves higher throughput, demonstrating that under the dynamic network environment the dynamic configurator performs co-inference with higher efficiency.
For example, set CDF as 0.6, the dynamic configurator makes 27 FPS throughput while the static configurator makes 17 FPS.
In addition, the CDF curve of dynamic configurator rises with 11 FPS throughput while the static configurator begins with 1 FPS, which indicates that the dynamic configurator works more efficiently than the static configurator at the beginning.

Fig. \ref{reward comparison} presents the CDF results of reward.
Similarly, under the same CDF level, the dynamic configurator acquires higher reward than the static configurator and the CDF curve of the dynamic configurator rises latter again.
However, in Fig. \ref{reward comparison} the two curves is closer than in Fig. \ref{conf comparison}, which means that the two configurators achieve nearly the same good performance from the perspective of reward.
This is because the two configurators make similar choices in the selection of exit point (i.e., in most cases both of them select exit point 5 as part of the co-inference strategy).
Therefore the difference of the reward mainly comes from the throughput result.
It demonstrates that the static configurator may perform as well as the dynamic configurator in some cases but the dynamic configurator is better in general under the dynamic network environment.

\section{Conclusion} \label{conclu}
In this work, we propose \textsf{Edgent}, an on-demand DNN co-inference framework with device-edge collaboration. Enabling low-latency edge intelligence, \textsf{Edgent} introduces two design knobs to optimize the DNN inference latency: DNN partitioning that enables device-edge collaboration, and DNN right-sizing that leverages early-exit mechanism.
We introduce two configurators that are specially designed to figure out the collaboration strategy under static and dynamic bandwidth environments, respectively. Our prototype implementation and the experimental evaluation on Raspberry Pi shows the feasibility and effectiveness of \textsf{Edgent} towards low-latency edge intelligence.
For the future work, our proposed framework can be further combined with existing model compression techniques to accelerate DNN inference.
Besides, we can extend our framework to support multi-device application scenarios by designing efficient resource allocation algorithms.
We hope to stimulate more discussion and efforts in the society and fully realize the vision of edge intelligence.

\section*{Acknowledgment}
This work was supported in part by the National Science Foundation of China under Grant No. U1711265, 61972432 and 61802449, the Program for Guangdong Introducing Innovative and Entrepreneurial Teams under Grant No. 2017ZT07X355, the Pearl River Talent Recruitment Program under Grant No. 2017GC010465, the Guangdong Natural Science Funds under Grant No. 2018A030313032, and the Fundamental Research Funds for the Central Universities under Grant No. 17lgjc40.

\ifCLASSOPTIONcaptionsoff
  \newpage
\fi

\bibliography{reference}

\begin{IEEEbiography}[{\includegraphics[width=1in,height=1.25in,clip,keepaspectratio]{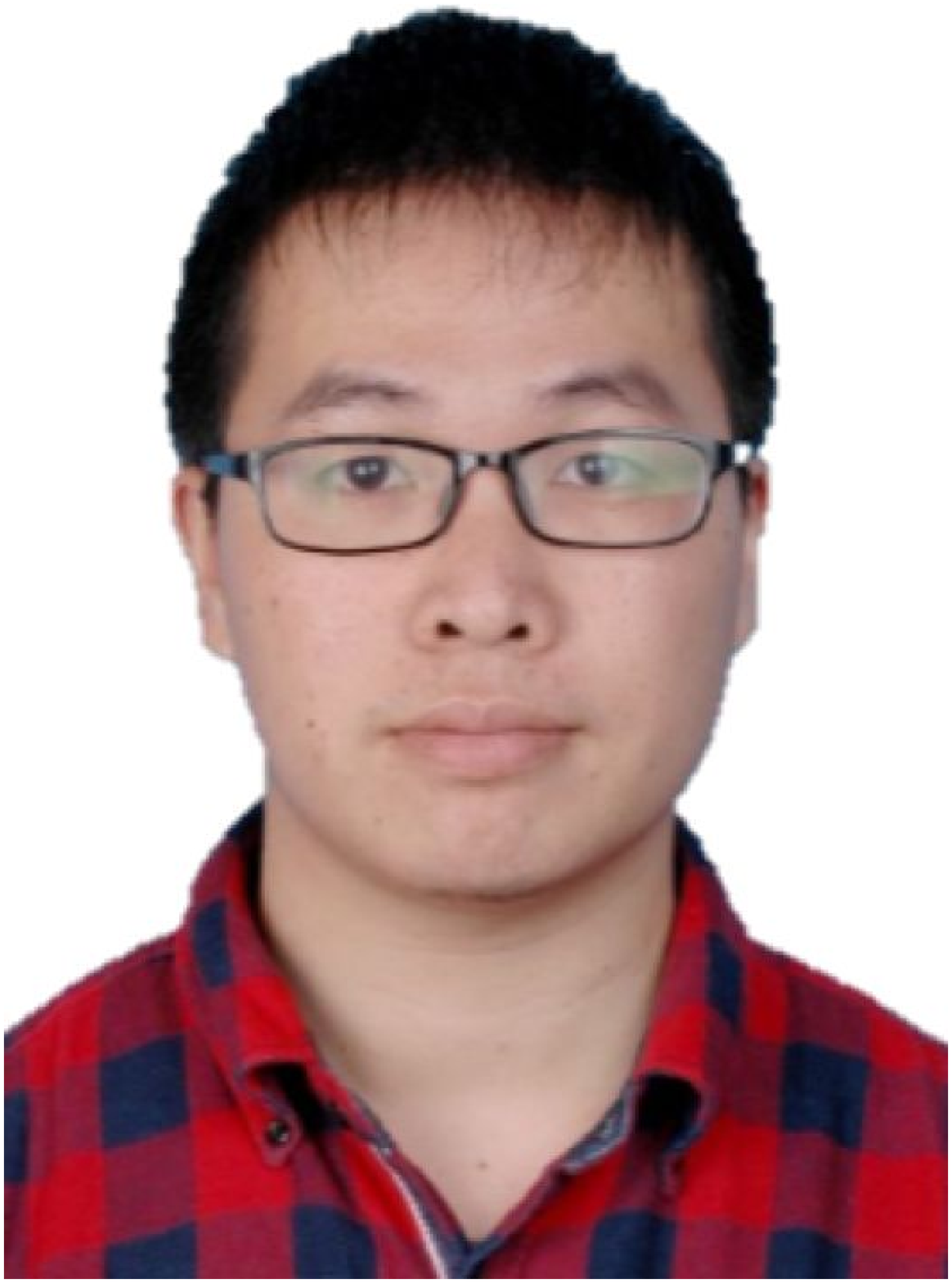}}]{En Li} received the B.S. degree in communication engineering from the School of Physics \& Telecommunication Engineering, South China Normal University (SCNU), Guangzhou, China in 2017, and M.S. degree in computer techonology from the School of Data and Computer Science, Sun Yat-sen University, Guangzhou, China. His research interests include mobile deep computing, edge intelligence, deep learning.
\end{IEEEbiography}

\begin{IEEEbiography}[{\includegraphics[width=1in,height=1.25in,clip,keepaspectratio]{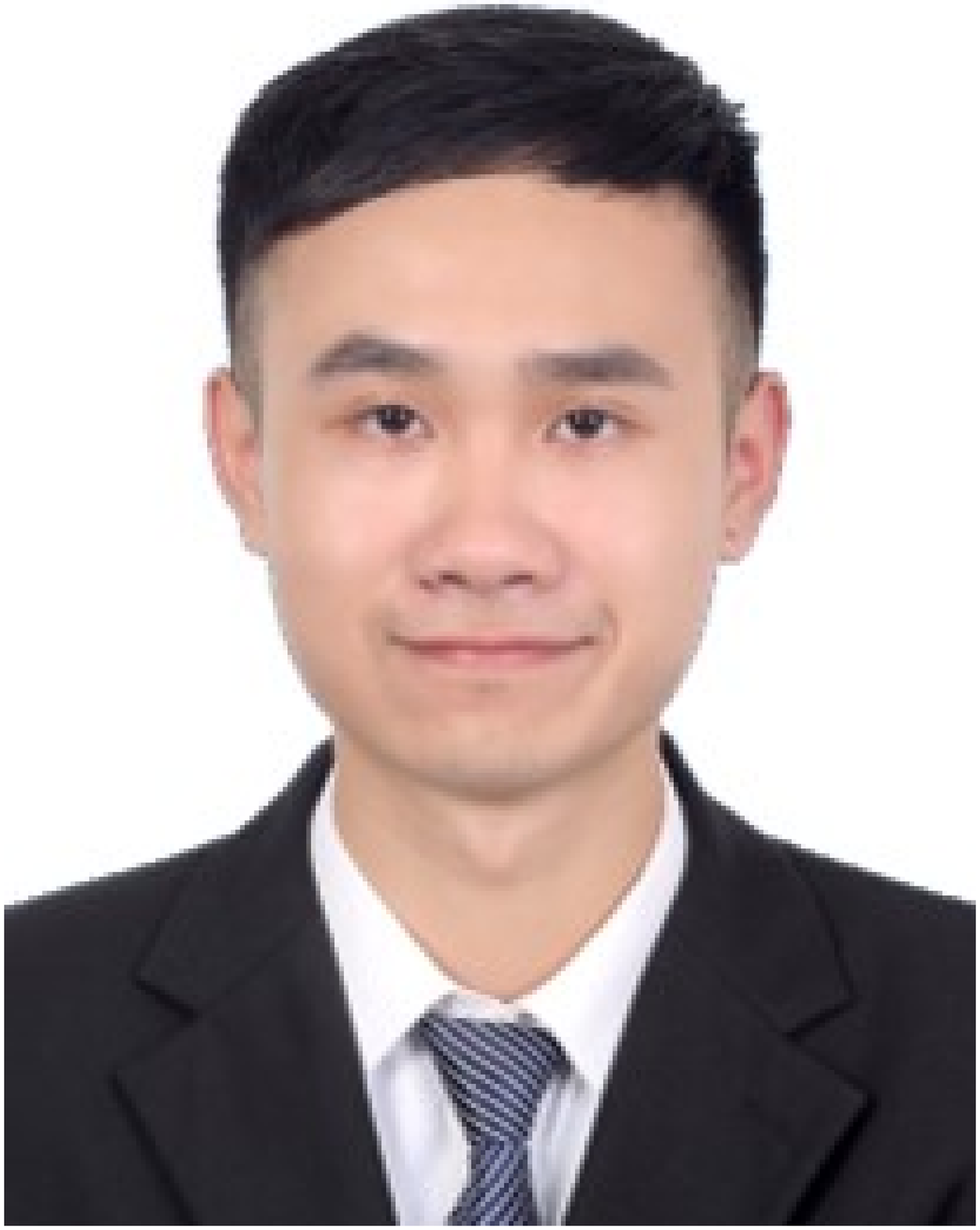}}]{Liekang Zeng} received the B.S. degree in computer science from the School of Data and Computer Science, Sun Yat-sen University (SYSU), Guangzhou, China in 2018. He is currently pursuing the master’s degree with the School of Data and Computer Science, Sun Yat-sen University, Guangzhou, China. His research interests include mobile edge computing, deep learning, distributed computing.
\end{IEEEbiography}

\begin{IEEEbiography}[{\includegraphics[width=1in,height=1.25in,clip,keepaspectratio]{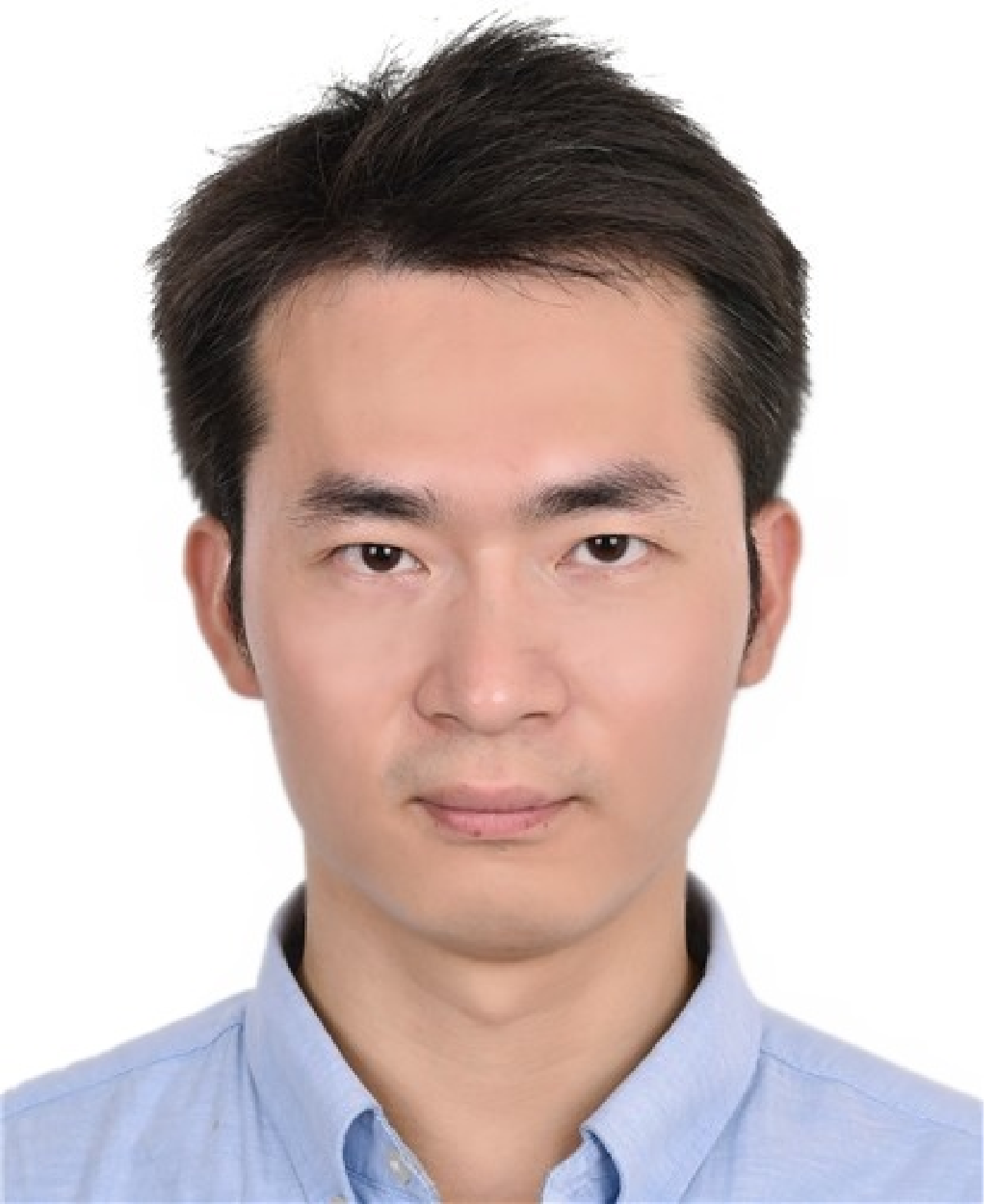}}]{Zhi Zhou}
received the B.S., M.E. and Ph.D. degrees in 2012, 2014 and 2017, respectively, all from the School of Computer Science and Technology, Huazhong University of Science and Technology (HUST), Wuhan, China. He is currently a research fellow in School of Data and Computer Science, Sun Yat-sen University, Guangzhou, China. In 2016, he has been a Visiting Scholar at University of Gottingen. He was the sole recipient of 2018 ACM Wuhan \& Hubei Computer Society Doctoral Dissertation Award, a recipient of the Best Paper Award of IEEE UIC 2018, and a general co-chair of 2018 International Workshop on Intelligent Cloud Computing and Networking (ICCN). His research interests include edge computing, cloud computing and distributed systems.
\end{IEEEbiography}

\begin{IEEEbiography}[{\includegraphics[width=1in,height=1.25in,clip,keepaspectratio]{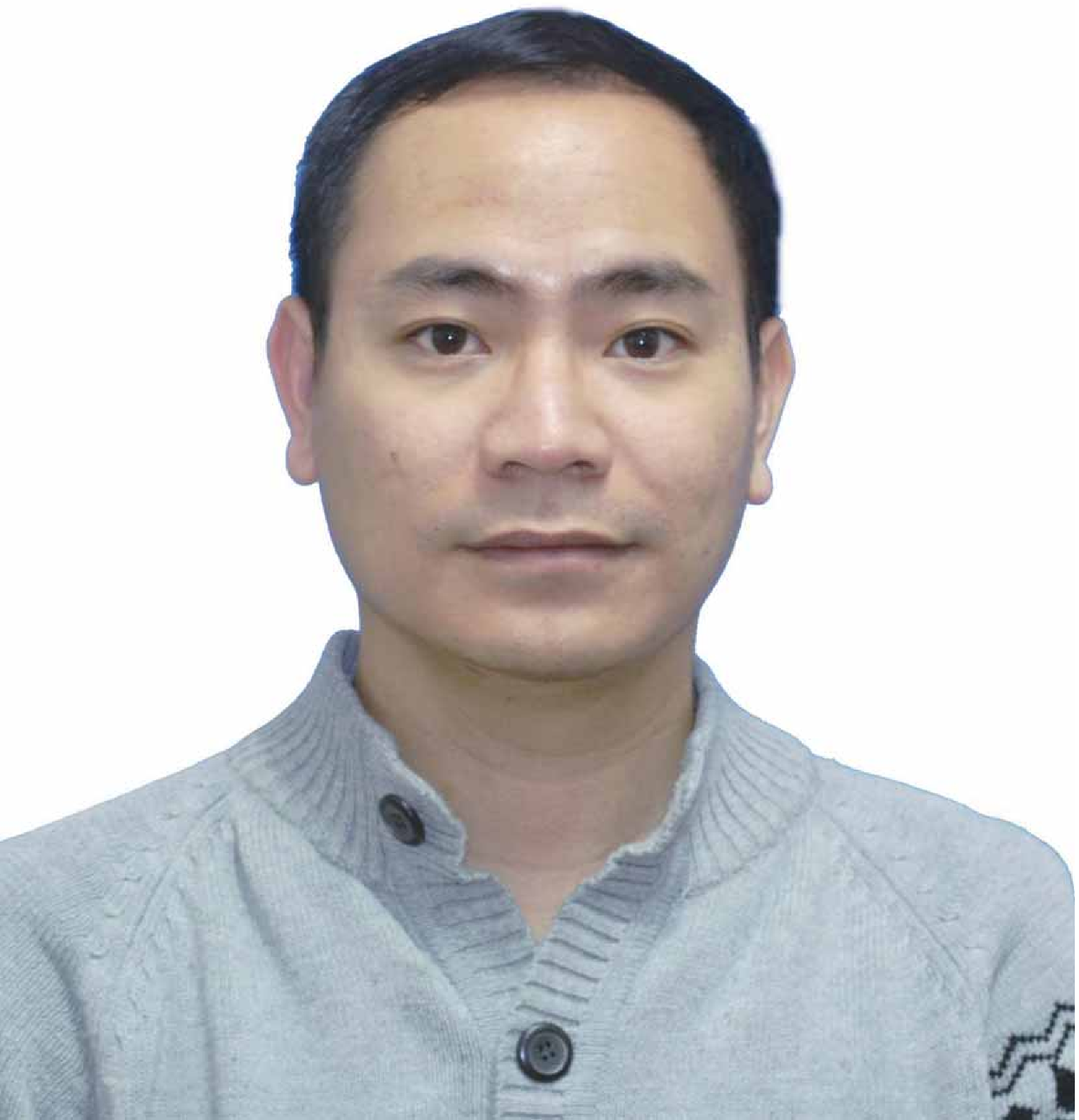}}]{Xu Chen}
is a Full Professor with Sun Yat-sen University, Guangzhou, China, and the vice director of National and Local Joint Engineering Laboratory of Digital Home Interactive Applications. He received the Ph.D. degree in information engineering from the Chinese University of Hong Kong in 2012, and worked as a Postdoctoral Research Associate at Arizona State University, Tempe, USA from 2012 to 2014, and a Humboldt Scholar Fellow at Institute of Computer Science of University of Goettingen, Germany from 2014 to 2016. He received the prestigious Humboldt research fellowship awarded by Alexander von Humboldt Foundation of Germany, 2014 Hong Kong Young Scientist Runner-up Award, 2016 Thousand Talents Plan Award for Young Professionals of China, 2017 IEEE Communication Society Asia-Pacific Outstanding Young Researcher Award, 2017 IEEE ComSoc Young Professional Best Paper Award, Honorable Mention Award of 2010 IEEE international conference on Intelligence and Security Informatics (ISI), Best Paper Runner-up Award of 2014 IEEE International Conference on Computer Communications (INFOCOM), and Best Paper Award of 2017 IEEE International Conference on Communications (ICC). He is currently an Area Editor of IEEE Open Journal of the Communications Society, an Associate Editor of the IEEE Transactions Wireless Communications, IEEE Internet of Things Journal and IEEE Journal on Selected Areas in Communications (JSAC) Series on Network Softwarization and Enablers.
 \end{IEEEbiography}

\end{document}